\documentclass[prb,aps,floatfix,twocolumn]{revtex4-1}%
\usepackage{amsfonts}
\usepackage{amsmath}
\usepackage{amssymb}
\usepackage{graphicx}%
\begin{document}
\preprint{HEP/123-qed}
\title[Charge Distribution Near Oxygen Vacancies in Reduced Ceria]{Charge Distribution Near Bulk Oxygen Vacancies in Cerium Oxides}
\author{E. Shoko, M. F. Smith, and Ross H. McKenzie}
\affiliation{University of Queensland, Department of Physics, Brisbane, QLD 4072, Australia}
\pacs{PACS number}

\begin{abstract}
Understanding the electronic charge distribution around
oxygen vacancies in transition metal and rare earth oxides is
a scientific challenge of considerable technological importance.
We show how significant information about the charge distribution around vacancies in cerium oxide can be gained from a study of
high resolution crystal structures of
 higher order oxides which exhibit ordering of oxygen vacancies. Specifically, we consider the implications
of a bond valence sum analysis of
 Ce$_{7}$O$_{12}$ and Ce$_{11}$O$_{20}$.
To illuminate our analysis we show alternative representations of the crystal structures in terms of orderly arrays of co-ordination defects and
in terms of flourite-type modules.
We found that in Ce$_{7}$O$_{12}$, the excess charge resulting
from removal of an oxygen atom delocalizes among all three triclinic Ce sites closest to the O vacancy. In Ce$_{11}$O$_{20}$, the charge localizes on the next nearest neighbour  Ce atoms. 
Our main result is that the charge prefers to distribute itself so that it is farthest away from the O vacancies. This contradicts \emph{the standard picture of charge localisation} which assumes that each of the two excess electrons localises on one of the cerium ions nearest to the vacancy. 
This standard picture is assumed in most calculations based on density functional theory (DFT).
Based on the known crystal structure of Pr$_{6}$O$_{11}$,
we also predict that the charge in Ce$_{6}$O$_{11}$ will be found in the second coordination shell of the O vacancy.
Although this review focuses on bulk cerium oxides our approach to characterising electronic properties of oxygen vacancies and the physical insights gained should also be relevant to surface defects and to other rare earth and transition metal oxides.
\end{abstract}
\volumeyear{year}
\volumenumber{number}
\issuenumber{number}
\eid{identifier}
\date[Date text]{date}
\received[Received text]{date}

\revised[Revised text]{date}

\accepted[Accepted text]{date}

\published[Published text]{date}

\startpage{01}
\endpage{102}
\maketitle

\tableofcontents


\section{Introduction}
\label{Sec I}
Non-stoichiometric phases of a number of transition metal and rare earth oxides 
have become widely studied in the past decade because of their potential
in technological applications. Important examples are derived from  the ``parent" materials TiO$_2$, ZrO$_2$, HfO$_2$, CeO$_2$, 
by doping with metal ions or reduction by removal of oxygen atoms.
A key scientific question in all these materials is the structure and
electronic properties of oxygen vacancies.
Ceria (cerium oxide) is a technologically important material with
applications in high temperature electrochemical devices \cite{Singhal2003,Inaba1996,Steele2000,Kharton2001,Zhang2006},
catalysis \cite{Trovarelli2002,Blank2007,Chen2007,Mo2005}, oxygen gas sensors \cite{Fray2001,Jasinski2003}, and magnetic semiconductors\cite{Keating2009,Song2009}. 
A fundamental property of this material relevant
to all of these application is its oxygen storage capacity (OSC). 
The material can rapidly take up and release oxygen
through a reversible chemical reaction.
There has been a considerable multi-disciplinary research effort aimed at
developing a fundamental picture of the microscopic processes associated
with this reversible chemical reaction.
Key questions include:
\begin{enumerate}
\item What is the \emph{origin} of the reversible uptake and release of
oxygen by ceria?
\item What is the \emph{origin} and \emph{mechanism} of the anionic
conduction?
\item What is the \emph{nature}, \emph{composition} and \emph{geometry} of
the catalytically active sites on cerium oxide surfaces?
\item What is the \emph{nature }of oxygen vacancies in the bulk solid and on
surfaces?
\item When an oxygen atom is removed to create a
vacancy, \emph{what happens to the two electrons that are left behind}?
\end{enumerate}
Similar questions are also important in other oxides.
This review is primarily concerned with
the last two questions and the study is restricted to the case of bulk
ceria. In this introduction, we briefly outline the current 
understanding of the problem with respect to the last two questions.

\begin{figure}
[ptb]
\begin{center}
\includegraphics[width=0.4\textwidth]
{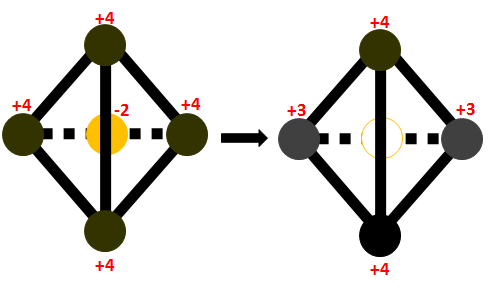}%
\caption{Schematic
of the standard picture of charge redistribution following the
formation of an oxygen vacancy in CeO$_{2}$. The tetrahedron of Ce atoms
(black circles) with an O atom at its centre (grey (orange in colour
version) circle) is shown along with the charges on these atoms in the
simple ionic picture description of CeO$_{2}$. The process of reduction
shown by the arrow leads to a neutral O vacancy at the centre of the tetrahedron
(empty circle) while two of the Ce ions have been
reduced to the +3 oxidation state.}%
\label{std}%
\end{center}
\end{figure}

When CeO$_{2}$ is reduced to the various defective phases, CeO$_{2-x}$,
according to Eq. \ref{reduce}, O vacancies are formed in the lattice
structure.
\begin{equation}
\text{CeO}_{2}\rightleftharpoons \text{CeO}_{2-x}+\frac{x}{2}\text{O}_{2}%
\text{ (g), \ \ \ \ }0\leq x\leq 0.5  \label{reduce}
\end{equation}%
The crystal structure adopted by any such defective phase, CeO$_{2-x}$, is
understood to be the one that provides the most favourable energetics for
the arrangement of all the O vacancies within the structure. In a widely
accepted view of the microscopic description of O vacancy formation and
ordering in CeO$_{2-x}$ phases, the two electrons associated with a missing
O atom when an O vacancy forms fully localize on two of the four equivalent
Ce$^{4+}$ ions which form a tetrahedron around the vacancy site as shown in
Eq. \ref{vacancy} \cite{Hoskins1995,Skorodumova2002,Trovarelli2002,Esch2005}:
\begin{equation}
4\text{Ce}_{\text{Ce}}+\text{ O}_{\text{O}}\rightleftharpoons 2\text{Ce}_{%
\text{Ce}}+\text{ }2\text{Ce}_{\text{Ce}}^{\prime }+V_{O}^{\cdot \cdot }+%
\frac{1}{2}\text{O}_{2}\text{\ (g)}  \label{vacancy}
\end{equation}%
where we have used the Kroger-Vink notation\cite{Kroger1956} so that the
symbols have the following meanings: Ce$_{\text{Ce}}$ - a Ce$^{4+}$ ion on a
Ce lattice site, O$_{\text{O}}$ - O$^{2-}$ ion on an O lattice site, Ce$_{%
\text{Ce}}^{\prime }$ - a Ce$^{3+}$ ion on a Ce lattice site and $%
V_{O}^{\cdot \cdot }$ - neutral O vacancy site.
This mechanism, which we will
henceforth refer to as the standard picture, is illustrated in the schematic
in Fig. \ref{std}.
%
The localization of an electron on a Ce$^{4+}$ ion converts it to the
slightly larger Ce$^{3+}$ ion with one electron in the $4f$ orbital. In the
reverse process where a defective phase, CeO$_{2-x}$, is oxidized, two $4f$
electrons from the two neighbouring Ce$^{3+}$ ion sites move onto the site
where an O atom is incorporated and then delocalize into the O $2p$ valence
band. Thus, the reversible processes of oxidation and reduction in Eq. \ref%
{reduce} have been considered to involve the extremal states of the Ce $4f$
electrons in which they are fully delocalized and fully localized
respectively \cite{Skorodumova2002}.

There have been, broadly speaking, three different approaches for
investigating this problem. Crystallographers have attempted to establish
general principles for O vacancy ordering in the reduced ceria phases. In
1974, Martin proposed the coordination defect model \cite{Martin1974}. He
suggested that when an O vacancy is created in the lattice of a rare-earth
oxide, the local environment of the vacancy undergoes relaxations in such a
way that the four nearest neighbour Ce and six nearest neighbour
O atoms form
a stable structural entity which is referred to as a coordination defect.
The crystal structures of the higher rare-earth oxides are then determined
and restricted by the topology of this coordination defect. In 1996, Kang
and Eyring developed a different framework for describing vacancy ordering
based on structural elements which are derived from the conventional unit
cell of CeO$_{2}$ in a simple way \cite{Kang1996}. As will be described in
more detail later, there are thirteen of these structural elements, called
modules, and they are the fundamental building blocks for all the crystal
structures of the higher oxides in the Kang-Eyring fluorite-type module
theory. As a result, it is possible to define what has been called a modular
unit cell in which, instead of the usual atomic basis, the different modules
form the basis. The value of this approach is the simplicity and elegance
with which it provides insight into the superstructures observed in the
reduced higher oxides in relation to the parent fluorite structure from
which they are derived.

Neither crystallographic approach sought to explicitly account for the
charge redistribution that occurs when an O vacancy is formed. However, as
we will show, they provide useful vantage points from which to examine the
problem. In contrast, the remaining two approaches in the strategy sought to
specifically establish the nature and occupation of the Ce $4f$ level in the
Ce oxides and consequently, the charge redistribution that occurs when an
oxygen vacancy is created. The earliest literature in this direction emerged
from spectroscopists \cite{Thornton1981} with their results being
interpreted using either cluster models \cite{Fujimori1983} or the single
impurity Anderson model \cite{Kotani1985,Kotani1988,Kotani2005}. This work investigated the
occupation of the Ce $4f$ level in the two extremal phases, CeO$_{2}$ and Ce$%
_{2}$O$_{3}$ and it was concluded that CeO$_{2}$ is mixed valence whereas Ce$%
_{2}$O$_{3}$ is a pure $4f^{1}$ configuration. The result that CeO$_{2}$ is
mixed valence is inconsistent with the ionic description underlying the
standard picture. However, the models used to interpret the spectroscopic
results are not parameter-free and the conclusions drawn from them have been
contested \cite{Sham2005}.

Recently electronic structure calculations based 
of density functional theory (DFT) have been performed. In general, DFT
results for CeO$_{2}$ assume empty Ce $4f$ $\ $states which is the simple
ionic picture of bonding in this oxide, e.g., Ref. \cite{Skorodumova2001}. To describe vacancy formation in CeO$%
_{2}$, the simplest case to consider is 
a $2\times 2\times 2$ supercell of CeO$_{2}$ from which a single O atom
is removed. This supercell has the composition Ce$_{32}$O$_{63}$ (CeO$_{1.97}$). As this
composition does not correspond to any known phases of reduced ceria, it is
customary to require the structural parameters of the relaxed structure to
match those of CeO$_{2}$ based on Vergard's rule \cite{Mogensen2000}, i.e., there is a linear relationship between the lattice constant and the extent of reduction. As will be discussed in more detail in Section \ref{comparison}, most of the DFT work supported the standard picture. However, the approximate functionals of DFT
such as LDA, GGA, and LDA+$U$ do not appear to describe well the electronic properties of the reduced phases \cite%
{Fabris2005,Ganduglia-Pirovano2007,Castleton2007}.
Furthermore, recent calculations using hybrid functionals
 \cite{ganduglia-pirovano:026101}
 and DFT+$U$ \cite{Castleton2007,ganduglia-pirovano:026101,li:193401} do not appear to support the standard picture.

As far as we know, a study of the charge distribution near bulk O vacancies in crystallographic phases of reduced cerium oxide
apart from Ce$_{2}$O$_{3}$ (and the customary supercell of DFT
calculations, i.e., CeO$_{1.97}$) has not been done. Our study of
the charge distribution in the well-characterized intermediate phases 
represents a complementary approach to answering
the fundamental question of where the two
electrons left behind when an O vacancy forms go. 
Furthermore, since no
in-situ studies have been done to identify the precise phases involved in
ceria-based catalysis, it is also possible that the intermediate phases may
play an important role in the engineering applications of these materials.

In this study, we examine the standard picture for the oxidation and
reduction in ceria described above, considering primarily two intermediate
phases namely Ce$_{11}$O$_{20}$ and Ce$_{7}$O$_{12}$. A brief discussion of
Ce$_{6}$O$_{11}$ is also included. We discuss this in light of our recent
results on Ce site valencies obtained by an analysis of observed crystal
structures using the bond valence method \cite{Shoko2009}.
  The two
cases, Ce$_{11}$O$_{20}$ and Ce$_{7}$O$_{12}$, were chosen to illustrate
the evolution of the charge distribution around O vacancies that occurs with
reduction. Furthermore, apart from the extremal structures, namely CeO$_{2}$
and Ce$_{2}$O$_{3}$, these two examples along with Ce$_{3}$O$_{5}$ were the
only crystal structures of the Ce oxides for which we were able to obtain
highly accurate crystallographic data to enable the type of analysis we do
here. 

As is discussed later, both Ce$_{11}$O$_{20}$ and Ce$_{7}$O$_{12}$
have fluorite-related crystal structures. According to Kang and Eyring,
 Ce$_{7}$O$_{12}$ is the limiting intermediate oxide for which any further
reduction leads to loss of the fluorite structure of CeO$_{2}$ \cite%
{Kang1997}. However, the fluorite-related Ce$_{3}$O$_{5}$ phase was recently
observed in the phase diagram of the Ce-O system\cite{Zinkevich2006}. In
addition to the conventional unit cells, we will generally use the Kang and
Eyring structural principle of modular units to discuss O vacancy ordering
in reduced ceria phases \cite{Kang1997,Kang1996,Kang1998,Lopez-Cartes1999}. There will be occasional references to Martin's
coordination defect model \cite{Martin1974,Hoskins1995,Bevan2008} where this model may provide a better conceptual framework.

This paper is organized as follows: In Section \ref{Sec II} we present the
crystal structures of Ce$_{11}$O$_{20}$ and Ce$_{7}$O$_{12}$ both in the
conventional way as well as in the representation of the Kang-Eyring
fluorite-type module theory. We show how the Kang-Eyring fluorite-type
module theory helps to understand the structure of O vacancies in Ce$_{11}$O$%
_{20}$ and Ce$_{7}$O$_{12}$ in relation to the parent fluorite structure. In
Section \ref{Sec III}, we consider the implications of our   bond
valence calculations for the charge distribution in the local
environment of the O vacancies. Section \ref{Sec IV} discusses   how
our results conflict with the standard picture of charge localization in
reduced ceria phases. We then make a prediction of the charge distribution to be expected in the Ce$_{6}$O$_{11}$ crystal based on the bond valence model in Section \ref{predict}. The possibility of direct $f$-$f$ coupling between neighbouring Ce sites in crystals of the Ce oxides. We assess the validity of this idea from the Harrison method of universal parameters in Section \ref{matrix}. Conclusions are presented in Section \ref{Sec VII}.
%
\section{Alternative Representations of 
the Crystal Structures of Ce$_{11}$O$_{20}$ and Ce$_{7}$O$_{12}$}
\label{Sec II}
\subsection{Coordination defect model}

\begin{figure}
[ptb]
\begin{center}
\includegraphics[width=0.4\textwidth]
{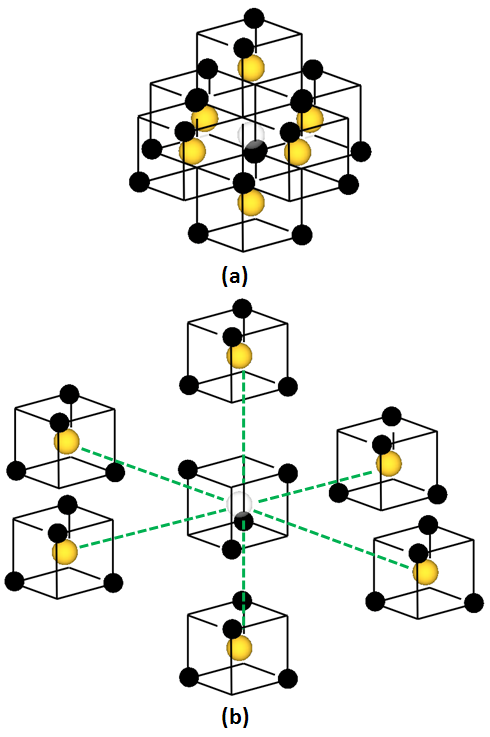}%
\caption{(a) Coordination defect showing
characteristic topology. (b) Exploded view of the coordination defect
showing the central octant of what is called the "$\protect\lambda $ phase"
coordinated by six M$_{\frac{1}{2}}$O octants of what is called the "$%
\protect\delta $ phase". The $\protect\lambda $ phase only differs from the $%
\protect\delta $ phase in that it has an O vacancy as shown. Black circles
represent Ce atoms, grey (orange in colour version) circles - O atoms and
the light grey circle inside the $\protect\lambda $ phase is an O vacancy.
After Martin \protect\cite{Martin1997}.}%
\label{coorddefect}%
\end{center}
\end{figure}

Martin proposed describing an O vacancy in a Ce oxide crystal as a
structural entity made up of $4$ Ce ions and $6$ O ions around the vacancy
site as illustrated in Fig. \ref{coorddefect} \cite{Martin1974,Hoskins1995}. The $4$ Ce ions form the first coordination shell (a
tetrahedron) of the vacancy site while the $6$ O atoms form the second
coordination shell (an octahedron). Compared to the perfect structure of CeO$%
_{2}$, all $4$ Ce ions are displaced $0.2 \text{\AA}$ away from the vacant O site while the O atoms are shifted $0.3 \text{\AA}$ towards the vacancy. This structural unit has been called the
coordination defect and it maintains its structural integrity in a crystal
of the oxide (see Fig. \ref{coorddefect}). It is constructed from two types of
subunits called octants, one of composition Ce$_{\frac{1}{2}}$O called the $%
\delta $ phase and the other incorporating an O vacancy, the $\lambda $
phase, of composition Ce$_{\frac{1}{2}}$. The exploded view in Fig. \ref%
{coorddefect} (b) shows these octants and how they are arranged to form the
coordination defect.
%
This model, in its original formulation, does not explicitly account for the
changes in the valences of the Ce ions and the four Ce ions which are
nearest neighbours to the O vacancy in the coordination defect are
implicitly identical. 

\subsection{Modules}

\begin{figure}
[ptb]
\begin{center}
\includegraphics[width=0.4\textwidth]
{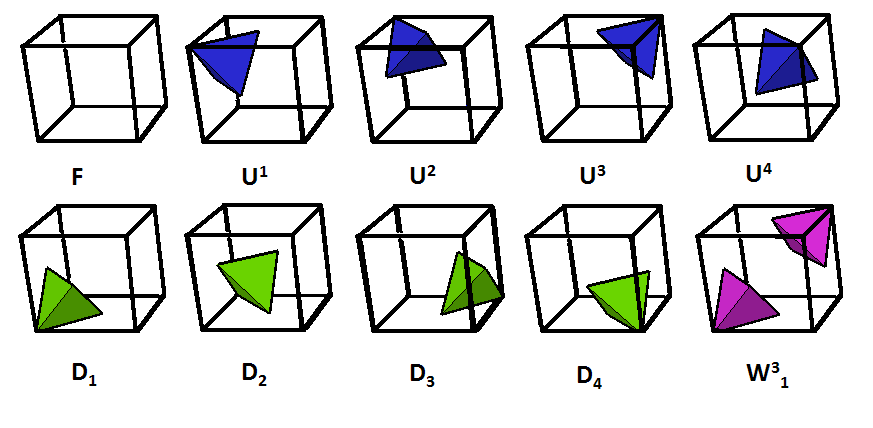}%
\caption{The ten different modules
required to build the modular unit cells of Ce$_{11}$O$_{20}$ and Ce$_{7}$O$%
_{12}$. The vertices of the tetrahedra denote the locations of the Ce ions
surrounding an O vacancy. Module $F$ is the conventional unit cell of CeO$%
_{2}$. Except for $W_{1}^{3}$ which has two O vacancies along the body
diagonal of the CeO$_{2}$ unit cell, the rest of the modules have only one O
vacancy in the positions shown. [Adapted from \protect\cite{Kang1997}]}%
\label{modules02}%
\end{center}
\end{figure}

Soon after the elaboration of Martin's coordination
defect model \cite{Hoskins1995}, Kang and Eyring developed a simpler model
for constructing the crystal structures of the higher rare earth oxides,
i.e., oxides from Ce$_{7}$O$_{12}$ to CeO$_{2}$ based
on a different set of structural elements \cite%
{Kang1996}. These  so-called modules, include the
CeO$_{2}$ unit cell \ with the rest derived from it by creating one or two O
vacancies in certain prescribed ways resulting in a total of thirteen
modules. These modules are the fundamental building blocks for all the
higher oxides under a prescription of rules defined by the authors \cite%
{Kang1997}. In Fig \ref{modules02}, we show the ten modules relevant to Ce$%
_{11}$O$_{20}$ and Ce$_{7}$O$_{12}$ in addition to the unit cell of CeO$_{2}$%
.
%
Kang and Eyring's method accounted for all the higher oxides known at the
time and was successful in predicting the existence of new structures \cite%
{Kang1997}. The Kang and Eyring method enables a relatively simple way of
visualizing vacancy sites which are otherwise difficult to decipher when the
structure is cast in the conventional setting of its space group. This
provides considerable facility in the analysis of the local environment of
the O vacancy site. As a result, we will make frequent reference to this
framework in the rest of this paper.
\subsection{Ce$_{11}$O$_{20}$}
\begin{figure}
[ptb]
\begin{center}
\includegraphics[width=0.4\textwidth]
{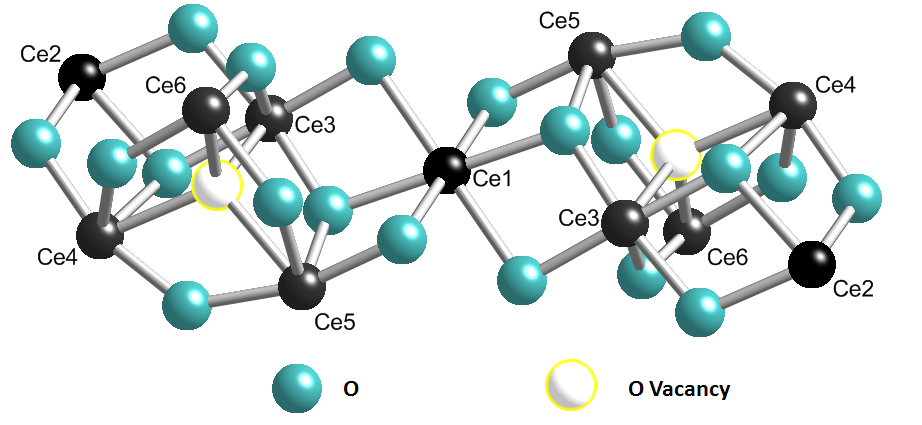}%
\caption{The proposed vacancy cluster of
Ce$_{11}$O$_{20}$ consisting of two O vacancies which are nearest neighbours
and related to each other by inversion symmetry centred on the Ce(1) site.
The various Ce sites are labelled as shown. The white and light grey (teal)
circles represent O vacancies and various O sites respectively.}%
\label{ce11o20divacancy}%
\end{center}
\end{figure}

The conventional unit cell of Ce$_{11}$O$_{20}$ is shown in Fig. \ref%
{ce11o20unitcell}. Note that for all ball and stick figures in this paper,
the sticks between the Ce sites and the O vacancy site do not represent
chemical bonds. Instead, the sticks are only included to show the geometric
relationships between the atoms. The unit cell of Ce$_{11}$O$_{20}$ consists
of one formula unit per unit cell with inversion as its point group
symmetry. It has six Ce and ten O distinct sites 
along with their inversion images with the Ce(1) site as the centre of
inversion. However, for a description of the vacancy structure and charge
distribution in this crystal, it is convenient to view all the lattice
points in the O sublattice as roughly equivalent and to divide the Ce
sublattice into two regions. The first region consists of the Ce(1) and
Ce(2) sites which, in the Ce sublattice, occupy the second coordination
shells of the O vacancies. The remainder of the Ce atoms are the first
coordination shells of the O vacancies.
\begin{figure}
[ptb]
\begin{center}
\includegraphics[width=0.4\textwidth]
{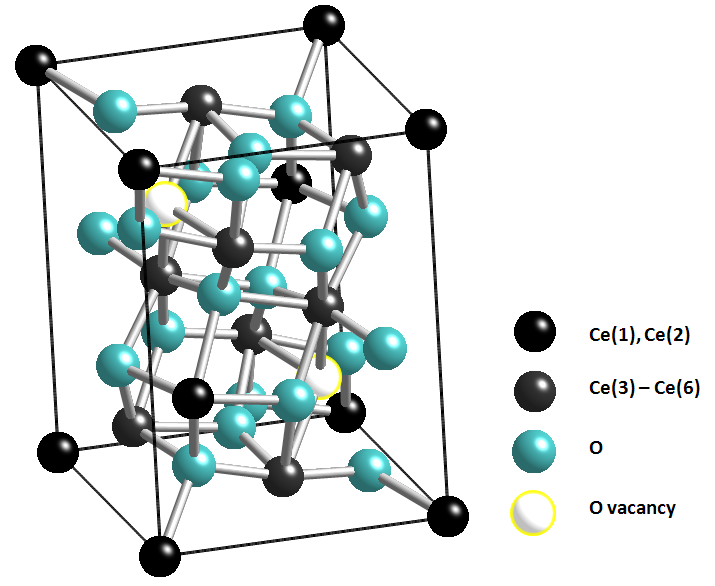}%
\caption{The unit cell of Ce$_{11}$O$%
_{20}$ whose space group is $P\bar{1}$. There are six distinct Ce sites in
this crystal. However, to facilitate the discussion of vacancy structures
and charge distribution, the Ce sublattice has been partitioned into two
types of lattice points as shown in the legend. The first set of lattice
points consists of the Ce(1) and Ce(2) sites which, if we only consider the
Ce sublattice, occupy the second coordination shells of the O vacancies. The
second set consists of the remainder of the Ce sites and these form the
first coordination shell of the O vacancies in the Ce sublattice of the
crystal. There are ten distinct O sites in the crystal but for our purposes,
it is convenient to treat all the O sites in the O sublattice as roughly
equivalent as indicated in the legend.}%
\label{ce11o20unitcell}%
\end{center}
\end{figure}
%
The crystal structure of Ce$_{11}$O$_{20}$ can be readily related to the
fluorite structure of CeO$_{2}$ from which it is derived by reduction. 
This relationship becomes clearer with the Kang-Eyring 
module representation.      
 Kang and Eyring \cite{Kang1997} have shown that the modular unit cell of Ce$_{11}$O$_{20}$ has
the modular composition $3F$, $4D$, $4U$ where the module types are as given
in Fig. \ref{modules02}. They also provided the modular sequences for the $%
[100]$ and $[010]$ directions in this crystal as $%
FFU^{3}U^{2}D_{3}D_{2}FU^{4}U^{1}D_{4}D_{1}$ and $%
FU^{3}D_{3}FU^{1}D_{1}FU^{2}D_{2}U^{4}D_{4}$ respectively. Using this
information along with their module juxtaposition rules, we derived the
modular sequence for the $[001]$ direction to be $%
D_{1}D_{2}FFFU^{4}U^{3}U^{1}U^{2}D_{4}D_{3}$. This then enabled us to
construct the $3D$ modular unit cell of Ce$_{11}$O$_{20}$ shown in Fig. \ref%
{ce11o20module}. We note that this unit cell is not unique.
\begin{figure}
[ptb]
\begin{center}
\includegraphics[width=0.4\textwidth]
{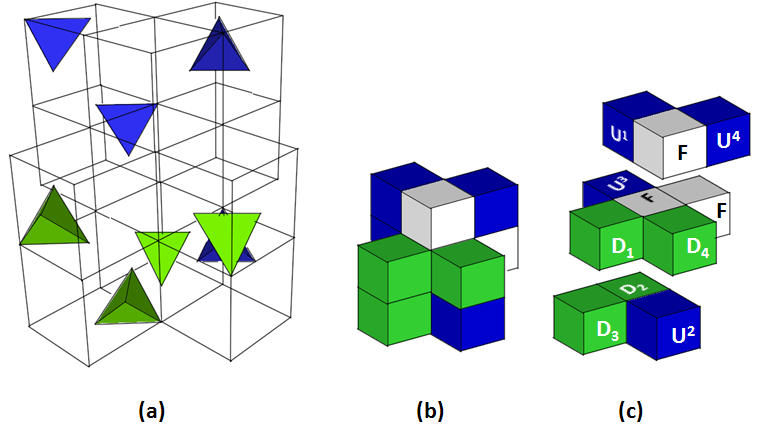}%
\caption{The fluorite-type modular unit
cell of Ce$_{11}$O$_{20}$ derived using the Kang and Eyring rules 
\protect\cite{Kang1997}. In (a), the modular unit cell is constructed from
transparent modules to show the location of the O vacancy site as indicated
by the shaded tetrahedra with the module types as identified in (c) and
shown in detail in Fig. \protect\ref{modules02}. A simplified block
representation of the modular unit cell given in (a) is shown in (b). The
exploded view with all the module types identified is given in (c). The
modular composition is: top layer - $U^{1}$, $F$, $U^{4}$; middle layer - $%
D_{1}$, $D_{4}$, $F$, $F$, $U^{3}$; bottom layer - $D_{2}$, $D_{3}$, $U^{2}$%
. These figures make clear the distribution of the O vacancies within the
crystal in relation to the original fluorite structure. There are large
contiguous regions, modules F, where there are no O vacancies.}%
\label{ce11o20module}%
\end{center}
\end{figure}
%
There is only one type of O vacancy in the crystal of Ce$_{11}$O$_{20}$ and
if we only consider the Ce sublattice, each O vacancy has the four
Ce(3)-Ce(6) sites forming its first coordination shell. These four Ce sites
form a distorted tetrahedron. The O vac - Ce($i$) distances are shown in
Table \ref{shells01} where $i=3,4,5,6$. The second coordination shell
consists of twelve Ce sites namely, two Ce(1), four Ce(2), two Ce(3), two
Ce(4) and one of each of the Ce(5) and Ce(6) sites as shown in Table \ref%
{shells01}. Of these twelve Ce sites in the O vacancy's second coordination
shell, only half a Ce(1) site and one Ce(2) site can be assigned to the
particular O vacancy site. Thus, in total, an O vacancy in Ce$_{11}$O$_{20}$
is proportionately associated with five and half Ce sites. It is convenient
to work with an integral number of Ce sites when studying the charge
distribution in the local environment of an O vacancy. In order to do this,
we propose to partition the lattice so that we consider a vacancy cluster
consisting of two neighbouring O vacancies which can then be associated with
eleven Ce sites. Since the choice of the Ce(1) site is the most difficult to
establish, we pick a Ce(1) site and construct the vacancy cluster around it.
A Ce(1) site has four closest O vacancies all in the third coordination
shell. Two of the O vacancies are located $4.558 \text{\AA}$ away from the Ce(1) lattice point with the other two at $4.571 \text{\AA}$. Since the Ce(1) site is a centre of inversion, each of these pairs of O
vacancies are related by inversion symmetry. We choose the pair with the
shortest distance from the Ce(1) site to define the vacancy cluster
consisting of two O vacancies. The cluster obtained by this procedure is
illustrated in Fig. \ref{ce11o20divacancy}.
%
This cluster should not be confused with the divacancy of Ce$_{7}$O$_{12}$
to be discussed later. The divacancy of Ce$_{7}$O$_{12}$ is a unique
structural entity of the crystal arising from long-range ordering of the
vacancies whereas the cluster we define here is simply an analytic
convenience. It cannot~be strictly considered to be specified in a unique
way since the pair of O vacancies we chose is only $0.01 \text{\AA}$ closer to the Ce(1) compared to the other. On the O vacancy sublattice,
each O vacancy has four nearest neighbours $6.134 \text{\AA}$ away for the first pair and $6.138 \text{\AA}$ for the second. There is no difficulty in identifying the Ce(2) site
associated with each O vacancy since there is only one closest ($4.495 \text{\AA}$) Ce(2) site. The distances are calculated from the crystallographic data of
Kummerle and Heger \cite{Kummerle1999}.
\begin{table}[tbp] 
\centering%
\caption[The first- and second coordination shells of the O vacancy]{The first- and second coordination shells of the O vacancy in the
Ce sublattice of Ce$_{11}$O$_{20}$ and some selected distances between the sites. These Ce-Ce distances show that the average intra-tetrahedral separation between Ce sites ($4.2\text{\AA}$) is larger than the inter-tetrahedral separation ($3.9\text{\AA}$).}%
\begin{tabular}{p{2cm} p{2cm} p{2cm} p{2cm}}
\hline
Ce site & Distance from O vacancy, $\text{\AA}$ & Direct Ce-Ce type & Direct Ce-Ce distance, $\text{\AA}$ \\ \hline
\multicolumn{4}{c}{First Coordination} \\ 
Ce(6) & $2.5106$ & Ce(6)-Ce(3) & $4.169$ \\ 
Ce(3) & $2.5674$ & Ce(6)-Ce(4) & $4.194$ \\ 
Ce(4) & $2.5817$ & Ce(6)-Ce(5) & $4.207$ \\ 
Ce(5) & $2.6185$ & Ce(3)-Ce(4) & $4.219$ \\ 
&  & Ce(3)-Ce(5) & $4.144$ \\ 
&  & Ce(4)-Ce(5) & $4.234$ \\ 
\multicolumn{4}{c}{Second Coordination} \\ 
Ce(5) & $4.3184$ & Ce(1)-Ce(5) & $3.810$ \\ 
Ce(4) & $4.4140$ & Ce(1)-Ce(4) & $6.904$ \\ 
Ce(3) & $4.4164$ & Ce(1)-Ce(3) & $3.852$ \\ 
Ce(3) & $4.4341$ &  &  \\ 
Ce(4) & $4.4390$ &  &  \\ 
Ce(6) & $4.4438$ & Ce(1)-Ce(6) & $5.561$ \\ 
Ce(2) & $4.4952$ & Ce(1)-Ce(2) & $7.715$ \\ 
Ce(2) & $4.5265$ & Ce(2)-Ce(3) & $3.873$ \\ 
Ce(2) & $4.5340$ & Ce(2)-Ce(4) & $3.861$ \\ 
Ce(1) & $4.5579$ & Ce(2)-Ce(5) & $6.860$ \\ 
Ce(1) & $4.5705$ & Ce(2)-Ce(6) & $5.573$ \\ 
Ce(2) & $4.6007$ &  &  \\ \hline
\end{tabular}%
\label{shells01}%
\end{table}%
Zhang \textit{et al}. \cite{Zhang1993} suggested another useful way of
viewing the crystal structure of Tb$_{11}$O$_{20}$ which is also directly
applicable to Ce$_{11}$O$_{20}$ because these compounds are isostructural.
Their perspective is based on the observation that the O vacancies are more
or less uniformly distributed within the crystal structure with no evidence
of vacancy pairing. The vacant O sites are distributed in a way such that
the separation between the defects is maximized and thus, the repulsive
interactions are reduced. The Ce ions in the first coordination shell relax
outward in a more or less isotropic way (Ce-$V_{O}^{\cdot \cdot }$ ranges $%
2.51-2.62 \text{\AA}$, see Table \ref{shells01}). As the vacant Ce tetrahedra expand, the
regions between them experience substantial compression as indicated by the
difference in the distances between the intra-tetrahedral ions ($4.2 \text{\AA}$) and those of the inter-tetrahedral ions ($3.9 \text{\AA}$). These distances are given in Table \ref{shells01}. In this table, the
intra-tetrahedral ion distances are given as the direct Ce-Ce distances in
the first coordination shell while the inter-tetrahedral ion distances refer
to these distances in the second coordination shell: Ce(1)-Ce(5),
Ce(1)-Ce(3); Ce(2)-Ce(3) and Ce(2)-Ce(4). The inter-tetrahedral Ce ions have
a coordination number of $8$ as in the parent fluorite structure. In
discussing the charge distribution in the local environment of the vacancy,
we will notice that the charge is distributed in a manner which is
distinctly different between the intra-tetrahedral and inter-tetrahedral Ce
ions.
\subsection{Ce$_{7}$O$_{12}$}
\begin{figure}
[ptb]
\begin{center}
\includegraphics[width=0.4\textwidth]
{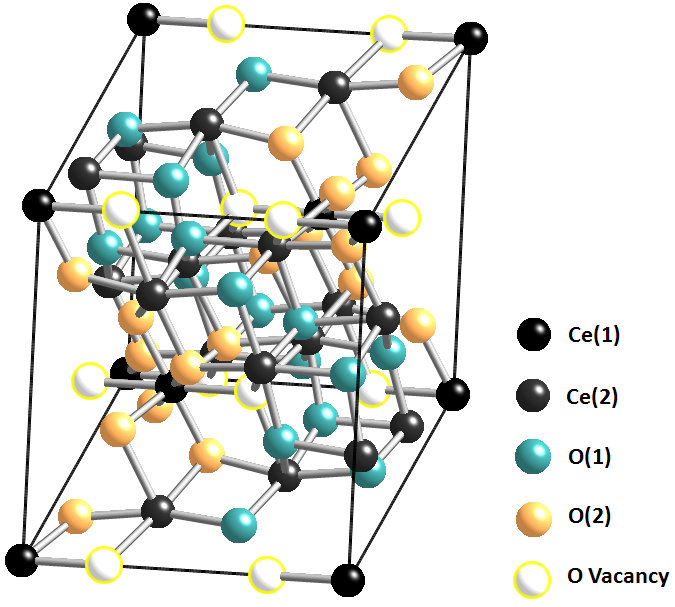}%
\caption{The unit cell of Ce$_{7}$O$%
_{12} $ showing the two distinct Ce sites: Ce(1) - black and Ce(2) - dark
grey. The O(1) sites are shown in grey (teal) with the O(2) light grey
(orange). The Ce(1) and Ce(2) sites are of $S_{6}$ and $i$ symmetries,
respectively. The O vacancy sites are shown in white. A characteristic
structural unit of the Ce$_{7}$O$_{12}$ crystal is the divacancy, which
consists of two O vacancy sites connected by an $S_{6}$ Ce site between
them.}%
\label{ce7o12unitcell}%
\end{center}
\end{figure}

\begin{figure}
[ptb]
\begin{center}
\includegraphics[width=0.4\textwidth]
{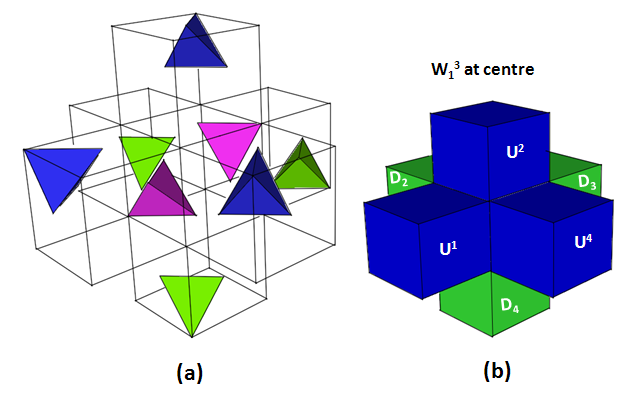}%
\caption{(a) The modular unit cell of Ce$%
_{7}$O$_{12}$ showing the details of the seven modules used to construct it.
Each tetrahedron represents the four Ce atoms around the oxygen vacancy.
(b) A simplified model of the modular unit cell. Adapted from Kang and
Eyring \protect\cite{Kang1997}.}%
\label{ce7o12module}%
\end{center}
\end{figure}

Fig. \ref{ce7o12unitcell} shows the unit cell of Ce$_{7}$O$_{12}$ in the
conventional rhombohedral setting. In the Kang-Eyring framework, the modular
composition of this crystal is $D_{2}$, $D_{3}$, $D_{4}$, $W_{1}^{3}$, $%
U^{1} $, $U^{2}$ and $U^{4}$ \cite{Kang1997}. The modular unit cell constructed from this
basis is depicted in Fig. \ref{ce7o12module}.
%
%
It is readily apparent from Fig. \ref{ce7o12module} where the O vacancy
sites are located. There are two types of Ce sites, Ce(1) and Ce(2), in the
crystal lattice of Ce$_{7}$O$_{12}$ which are of $S_{6}$ and $i$ symmetry
respectively \cite{Ray1975,Kummerle1999}. These distinct sites are
indicated in the conventional unit cell of Fig. \ref{ce7o12unitcell}.
Although Kang and Eyring did not explicitly identify the distinct sites in
their modular unit cell according to their symmetries, it is apparent that
by translating their unit cell in three dimensions, two distinct Ce sites
are obtained; the $S_{6}$ site with two nearest neighbour O vacancies around
it and the $i$ site with only one O vacancy. The $S_{6}$ site has been
called the divacancy site \cite{Kang1997} and it forms a shared corner
between two coordination tetrahedra of the O vacancies as illustrated in
Fig. \ref{ce7o12charge}.
We list here some interatomic distances that we will use in our analysis of
the divacancy cluster: Ce(2)-Ce(2) $=4.09 \text{\AA}$, Ce(1)-Ce(2) $=4.11 \text{\AA}$, Ce(1)-O(2) $=2.19 \text{\AA}$ and for Ce(2)-O(1), there are two relevant bond lengths, namely $2.23%
 \text{\AA}$ and $2.32 \text{\AA}$. The Ce-Ce distances are direct distances between these sites, not
distances along the Ce-O bonds connecting them. It appears that the Ce
sublattice has not changed much relative to CeO$_{2}$.

\begin{figure}
[ptb]
\begin{center}
\includegraphics[width=0.4\textwidth]
{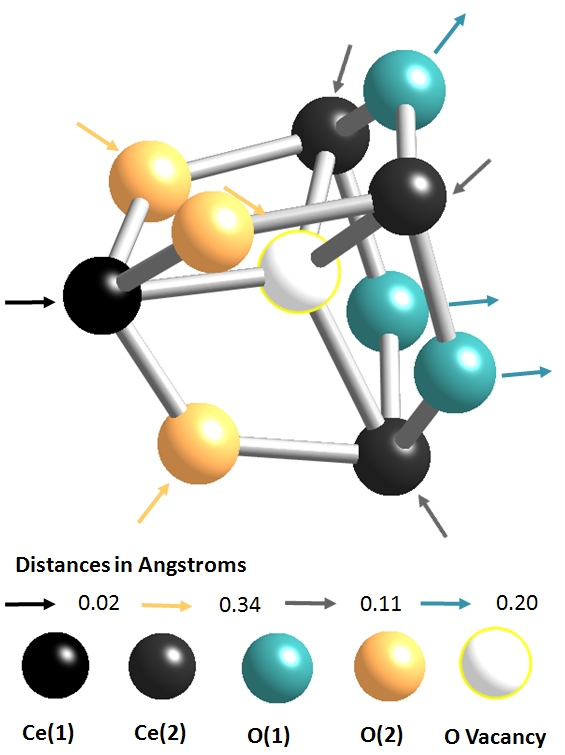}%
\caption{The relaxation of atoms forming
part of the divacancy in Ce$_{7}$O$_{12}$. The different lattice sites are
indicated by the colours as follows: Ce(1) - black, Ce(2) - dark grey, O(1)
- grey (teal), O(2) - light grey (orange) and the O vacancy ($V_{O}^{\cdot
\cdot }$) - white. The arrows indicate the direction in which the respective
atoms relax relative to the O vacancy site and the amounts by which the
`bonds' relax are shown in the legend. This figure was produced from the
crystal structure determined by neutron scattering \protect\cite%
{Kummerle1999}.}%
\label{relax}%
\end{center}
\end{figure}

It is helpful to visualize vacancy ordering in Ce$_{7}$O$_{12}$ by
viewing the divacancy as a structural unit of the O vacancies in this
crystal. This clarifies why the
ratio of $S_{6}$ to $i$ sites is $1:6$; all the Ce sites which are nearest
neighbours to the two O vacancies forming a divacancy are of $S_{6}$
symmetry. As there are seven Ce atoms in the formula unit, the rest of the
Ce atoms (six) must be the Ce(2) sites of $i$ symmetry. The formation of the
divacancy structures as CeO$_{2}$ is reduced to Ce$_{7}$O$_{12}$ is
associated with some relaxation of the crystal lattice although the Ce
sublattice remains somewhat invariant \cite{Kang1997}. We now examine this
relaxation process which is illustrated in Fig. \ref{relax} and based on
Kummerle and Heger's neutron diffraction data \cite{Kummerle1999}.
%
The bond lengths in CeO$_{2}$ are $2.434 \text{\AA}$ and $2.706 \text{\AA}$ for the Ce-O and O-O bonds respectively. When a divacancy is formed, the
bonds within the divacancy relax and diffraction data show that in Ce$_{7}$O$%
_{12}$ the `bond' lengths are \cite{Kummerle1999}: $2.505 \text{\AA}$ for O(1)-$V_{O}^{\cdot \cdot }$, $2.364 \text{\AA}$ for O(2)-$V_{O}^{\cdot \cdot }$, $2.418 \text{\AA}$ for Ce(1)-$V_{O}^{\cdot \cdot }$ and $2.543 \text{\AA}$ for Ce(2)-$V_{O}^{\cdot \cdot }$. This structural data indicates that
compared to CeO$_{2}$, the bonds within the divacancy have changed as
follows: the O(1)-$V_{O}^{\cdot \cdot }$ and O(2)-$V_{O}^{\cdot \cdot }$
`bonds' contract by $0.20 \text{\AA}$ and $0.34 \text{\AA}$ respectively while the Ce(2)-$V_{O}^{\cdot \cdot }$ `bond' gets longer by 
$0.109 \text{\AA}$. The Ce(1)-$V_{O}^{\cdot \cdot }$ `bond' is practically unchanged,
showing only a small contraction of about $0.02 \text{\AA}$. Thus, we see that, just as was the case for Ce$_{11}$O$_{20}$, the O
atoms are attracted to the vacancy site while the Ce(2) atoms are repelled
away so that the O atoms now become the nearest neighbours of the vacancy
site. This appears to be a general feature of vacancies in rare earth higher
oxides \cite{Hoskins1995,Kang1997}. We note that the six O atoms
around the vacancy relax by different amounts with the O(2) atoms being more
strongly attracted to the vacancy site. We also note that according to the
coordination defect model, the crystal structure of Ce$_{7}$O$_{12}$ has the
highest packing density of the coordination defect.
\section{Excess electrons delocalise away from the oxygen vacancy}
\label{Sec III}
Having discussed the O vacancy ordering in Ce$_{11}$O$_{20}$ and Ce$_{7}$O$%
_{12}$ as well as proposing some vacancy clusters suitable for the analysis
of charge distribution in these crystals in the previous section, we now
present the results of the bond valence calculations, \textbf{I,} in these
clusters.
\subsection{Ce$_{11}$O$_{20}$}
\begin{figure}
[ptb]
\begin{center}
\includegraphics[width=0.4\textwidth]
{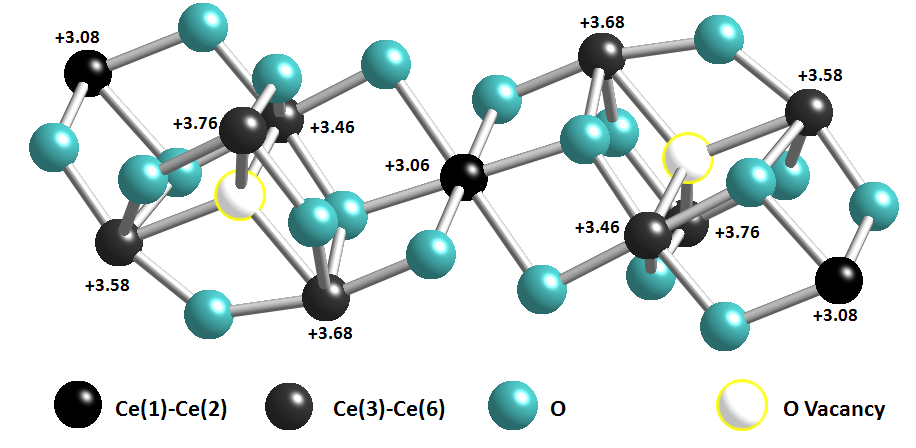}%
\caption{Charges  of the different Ce ions
in the vacancy cluster of Ce$_{11}$O$_{20}$ obtained from the bond valence
model. The vacancy cluster consists of two vacancies \ connected
by the Ce(1) site which is also a centre of inversion for the cluster.
The four excess electrons that arise due to the removal of the two oxygen
atoms do not localise on the cerium atoms nearest the vacancies. This is contrary to the standard picture.}%
\label{ce11o20divacancycharge}%
\end{center}
\end{figure}

Fig. \ref{ce11o20divacancycharge} shows the bond valence sums on the
various Ce sites in the vacancy cluster of Ce$_{11}$O$_{20}$ which was
introduced earlier, see Fig. \ref{ce11o20divacancy}. Clearly the charge
from the vacancy does not localize on the Ce sites closest to the vacancy
but rather, delocalizes onto the Ce(1) and Ce(2) sites.
This  contradicts the standard picture.
%
The results of Fig. \ref{ce11o20divacancycharge} can be compared with
those of Tb$_{11}$O$_{20}$ with which it is isostructural. Based on a simple
consideration of the crystal radii of the cations \cite{Shannon1976}, Zhang 
\textit{et al}. \cite{Zhang1993} suggested the following valencies for the
cations: Tb(1) and Tb(2) - $+3$, Tb(3) and Tb(4) - $+3.75$, Tb(5) and Tb(6)
- $+4$. Here the numbering of the sites is the same as for Ce$_{11}$O$_{20}$
given in Fig. \ref{ce11o20divacancy}. Their assigment of the site valences
qualitatively agrees well with our bond valence results for Ce$_{11}$O$_{20}$%
. The Tb$^{3+}$ and Tb$^{4+}$ ions are $f^{8}$ and $f^{7}$ configurations
respectively and can be compared to the corresponding $f^{1}$ and $f^{0}$
configurations respectively for the Ce ions.
They suggested the possibility of fast
 electron transfer between the neighbouring Tb(2)
sites so that, instead of two Tb(2)$^{3+}$ and one Tb(2)$^{4+}$
instantaneous states, an average is obtained from the neutron diffraction
data. It was suggested by the authors that the thermal ellipsoids of the
atoms in Ce$_{7}$O$_{12}$ which are much higher than in CeO$_{2}$ could be
evidence of the dynamic disorder arising from the fast electron transfer.

\subsection{Ce$_{7}$O$_{12}$}
\begin{figure}
[ptb]
\begin{center}
\includegraphics[width=0.4\textwidth]
{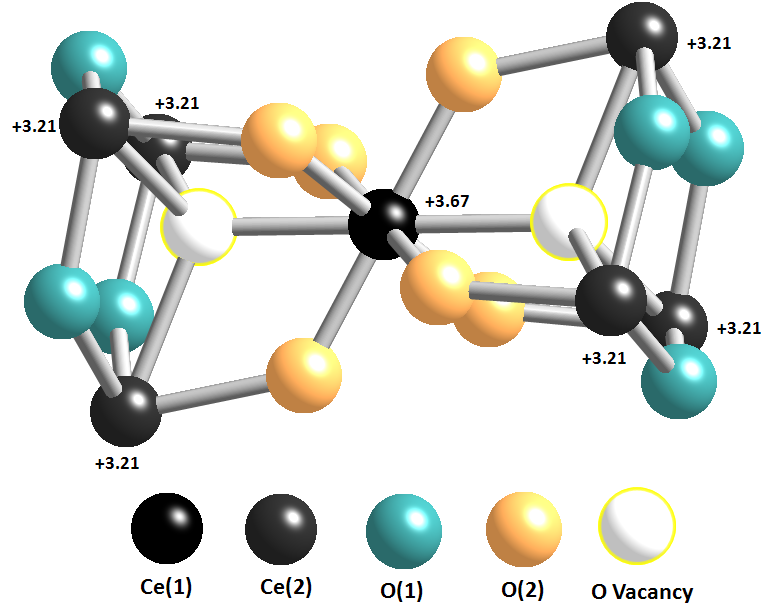}%
\caption{The divacancy of Ce$_{7}$O$%
_{12} $ showing the valences of the Ce atoms calculated from the bond
valence model, \textbf{I}. The charge delocalizes on the three Ce(1) sites
of triclinic symmetry which now have a valence of $+3.21$.}%
\label{ce7o12charge}%
\end{center}
\end{figure}

The charge distribution in the divacancy of Ce$_{7}$O$_{12}$ is shown in
Fig. \ref{ce7o12charge}. This figure clearly shows that, apart from the
Ce(1) site which is of a different local site symmetry compared to the Ce(2)
sites, the charge delocalizes over all three of the Ce atoms closest to the
vacancy.
Again, this  contradicts the standard picture, which has the electrons completely localise on two of the neighbouring Ce atoms.
%
\section{Failure of the standard picture of charge localisation near a vacancy}
\label{Sec IV}
We now discuss the above results comparing them to the standard picture and
propose an alternative view of how the charge redistributes itself following
O vacancy formation in ceria as deduced from the bond valence model. As
already mentioned, the conventional description of the electronic processes
involved during O vacancy formation requires that the excess charge at the
vacant O site be shared between two of the four Ce ions which form the first
coordination shell of the vacancy. This means that the tetrahedron formed by
the four Ce ions around the O vacancy consists of two Ce$^{3+}$ and two Ce$%
^{4+}$ ions. This description is based on the ionic model for these oxides
although Kang and Eyring do indicate that there may be some covalent
character in these oxides \cite{Kang1997}. 
%
\subsection{Ce$_{11}$O$_{20}$}
\begin{figure}
[ptb]
\begin{center}
\includegraphics[width=0.4\textwidth]
{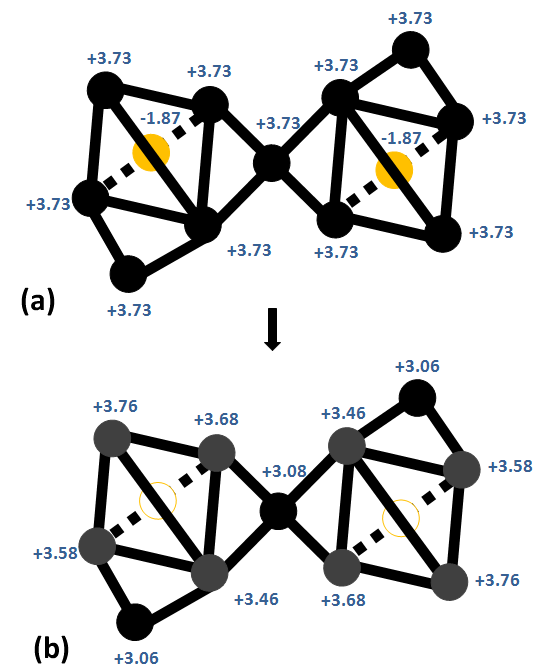}%
\caption{A schematic to illustrate the
formation of Ce$_{11}$O$_{20}$ from CeO$_{2}$ through vacancy formation in
the O sublattice. In (a), a part of the CeO$_{2}$ lattice, which, after
removing the two O atoms (light grey (orange) circles), gives a cluster
corresponding to the one given in Fig. \protect\ref{ce11o20divacancy}. Also
note that compared to Fig. \protect\ref{ce11o20divacancy}, here we have
made the further simplification of replacing all the Ce-O-Ce bonds with
direct Ce-Ce bonds except for the two O atoms which are subsequently removed
to create vacancies. For these two atoms, no bonding detail is given in the
schematic except that we place each of them at the centre of a regular Ce
tetrahedron reflecting their coordination environment in CeO$_{2}$. The
numbers shown at the atomic sites are the valences of the respective atoms
as determined by the bond valence method\protect\cite{Shoko2009},
and Fig. \protect\ref%
{ce11o20divacancycharge}. In (b), the cluster of (a) after reduction to
create the two O vacancies (empty circles). The colours in this figure
denote the following: black circles - Ce sites which are not nearest
neighbours to any O vacancies, grey circles - Ce atoms which are the nearest
neighbours of at least one O vacancy, light grey (orange) circles - O atoms
and empty circles - O vacancies.}%
\label{ce11o20bvm2}%
\end{center}
\end{figure}

If we assume that the four electrons which are left when the two O vacancies
are formed are localized within the cluster, then we can describe how the
charge redistributes itself. In fact, from the bond valence results,
which also showed mixed valence in CeO$_{2}$, the
valence of Ce in CeO$_{2}$ is $+3.73$ and the corresponding valence of O is $%
-1.87$. A description of vacancy formation in CeO$_{2}$ to form Ce$_{11}$O$%
_{20}$ can then be illustrated in the bond valence model as shown in Fig. %
\ref{ce11o20bvm2}.
%
Thus, according to the mixed valence description of the CeO$_{2}$
crystal from the bond valence model (see Fig. \ref{ce11o20bvm2}(a)), when
an O vacancy is created in CeO$_{2}$, a charge of $-1.87$ is left behind as
opposed to a charge of $-2.0$ in the ionic picture. This then means that, if
we consider the vacancy cluster of Fig. \ref{ce11o20divacancycharge} and
equivalently, Fig. \ref{ce11o20bvm2}(b), the creation of the two O
vacancies leaves a total charge of $-3.74$ in the cluster. From Fig. \ref%
{ce11o20divacancycharge}, we can see that compared to their original
valences in CeO$_{2}$, the valences of Ce(5) and Ce(6) are virtually
unchanged. Thus, considering the change in the valences of the remainder of
the Ce sites, the total charge accummulation on the Ce sites in the cluster
is $-2.87$. There is therefore a discrepancy of $\sim -1$ between the total
extra charge in the Ce$_{11}$O$_{20}$ crystal inferred from the bonding in
CeO$_{2}$ (i.e. $-3.74$) and that directly calculated for this crystal by
the bond valence calculations ($-2.87$). This discrepancy is entirely
consistent with the accuracy of the bond valence model reported in \textbf{I}
although we note that the accounting of charge we have done here from CeO$%
_{2}$ to Ce$_{11}$O$_{20}$ is very simplistic. It ignores the many-body
effects which may follow the injection of extra charge at the Ce sites when
the O vacancies are created. Such an extra charge may change the degree of
mixing between the Ce $4f$- and the O $2p$ states in manner that may not be
related in a simple way to their state in CeO$_{2}$. It may happen that, for
instance, due to a change in the hybridization of Ce $4f$- and the O $2p$
states, the extra charge may not be confined to the Ce $4f$ \ as we assumed
in calculating the discrepancy in the total charge, but rather, some of the
charge may go into the O valence band. It is expected that the results in
Fig. \ref{ce11o20divacancycharge} would capture these many-body effects.
An important point from the results of Fig. \ref{ce11o20divacancycharge}
is that $-1.99$ ($\sim 70\%$) of the total charge of $-2.87$ goes to the
three Ce atoms in the second coordination shell with the remainder being
shared among the eight Ce atoms in the first coordination shell.

\ If one just considers the cation sublattice, it is evident that the charge
accumulates in regions of high strain (inter-tetrahedral Ce ions) where the
Ce-Ce distances are slightly shorter compared to the intra-tetrahedral case.
This led some authors to consider the possibility that there may be some
direct $f$-$f$ hopping in these dense regions \cite{Zhang1993}. As we will
discuss in Section \ref{matrix} for the case of Ce$_{7}$O$_{12}$, direct $f$-%
$f$ hopping appears unlikely in these oxides.
\subsection{Ce$_{7}$O$_{12}$}
\begin{figure}
[ptb]
\begin{center}
\includegraphics[width=0.4\textwidth]
{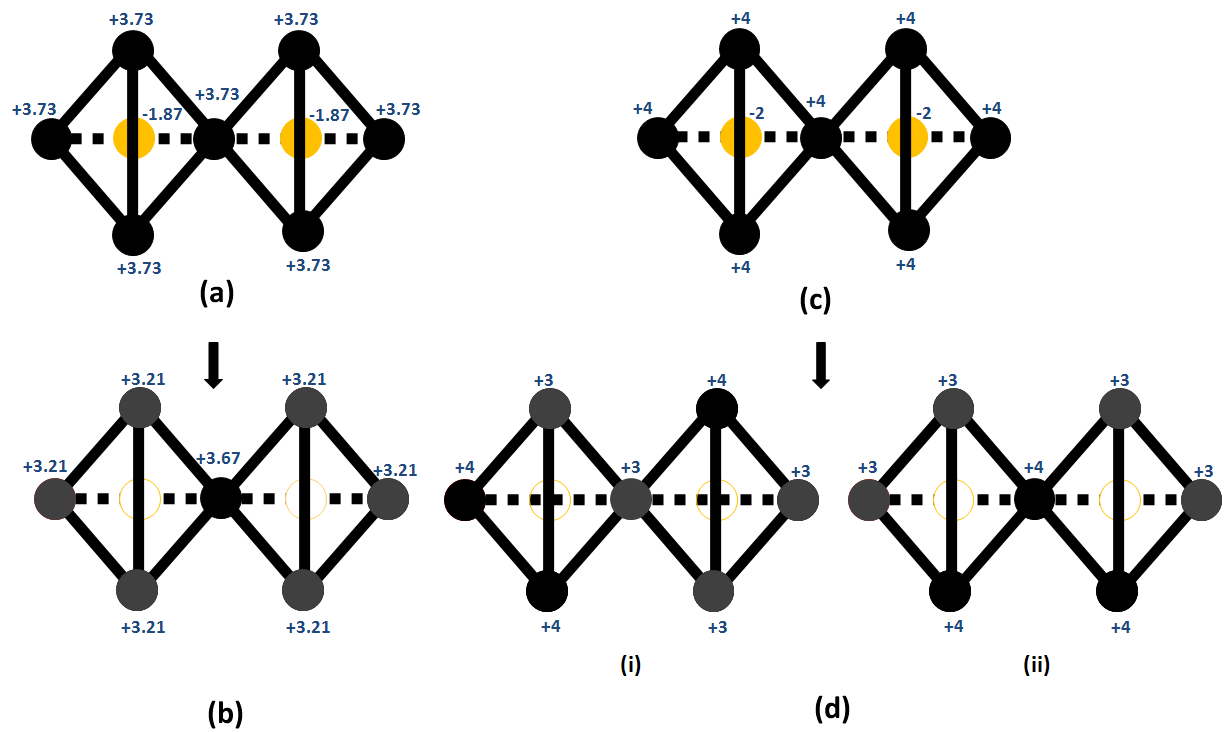}%
\caption{Schematic illustrating the charge distribution 
following divacancy formation,
based on a bond valence sum analysis of the crystal
structures of CeO$_{2}$
and Ce$_{7}$O$_{12}$.
(a) Part of
the CeO$_{2}$ lattice showing two neighbouring O atoms (light grey (orange)
circles), each inside a tetrahedron of Ce atoms (black circles). The numbers
shown are the bond valence sums for each of the atoms.
When the two neigbouring O atoms are removed due
to the reduction of CeO$_{2}$, the divacancy of Ce$_{7}$O$_{12}$ is formed
as shown in (b).
The bond valence sums on the Ce sites of
triclinic symmetry (grey circles) have decreased from $+3.73$ in CeO$_{2}$
to $+3.21$ reflecting the presence of the additional charge left behind when
the two O atoms are removed. The bond valence sum on the $S_{6}$ Ce site is
virtually unchanged.}%
\label{ce7o12bvm1}%
\end{center}
\end{figure}

Based on the results of Fig. \ref{ce7o12charge}, we can now discuss the
electronic features of the local environment of the O vacancy in Ce$_{7}$O$%
_{12}$. The formation of the Ce$_{7}$O$_{12}$ divacancy from the CeO$_{2}$
crystal lattice according to the BVM is schematically illustrated in Fig. \ref{ce7o12bvm1}(a) and (b). It can be seen that two corner-sharing Ce tetrahedra of CeO$_{2}$ lose an O
atom each from the centre of the tetrahedron. 
%
Our bond valence calculations show that the valences of the Ce(1) and Ce(2)
sites in Ce$_{7}$O$_{12}$ are $+3.67$ and $+3.21$ respectively. The valence
at the Ce(1) site is comparable to the value of $+3.73$ calculated for Ce in
CeO$_{2}$ using a bond valence method analysis of the crystal structure. If
we use the Kang and Eyring modular unit cell with its $8$ O vacancies, $4$
Ce(1) and $24$ Ce(2) sites and further assume that we start from
mixed-valence CeO$_{2}$ where the valences are $+3.73$ and $-1.87$ for the
Ce \ and O sites respectively, we can get some estimate of how the excess
charge on the O vacancy sites is distributed in the modular unit cell. The
valence of the Ce(1) sites does not change much from its value before the O
vacancy is created. Thus, 
 virtually all the charge left at the vacancy site by an O atom is
evenly distributed within the first coordination shell of the vacancy site
among the (three) Ce(2) sites. On this basis and only considering a simple
analysis using the bond valence results for CeO$_{2}$, the valence at the
Ce(2) sites is then readily calculated to be $+3.11$ (i.e., $3.73-1.87/3$)
which is comparable to the independently calculated value of $+3.21$ within
the error bounds ($\pm 0.1$) of the method. That the excess charge is not
distributed evenly among all four of the Ce atoms in the first coordination
shell contradicts the standard picture. 
If the charge was evenly distributed among all
four Ce atoms, the valences would be $+2.79$ and $+3.26$ for the $S_{6}$ and 
$i$ sites respectively.

The above picture from the BVM can be contrasted with standard picture description of the same process as illustrated in Fig. \ref{ce7o12bvm1}(c) and (d). Here, there are two ways one could assign the valences of the Ce sites in the divacancy. One way is to assign valences of $+4$ to three of the $i$ sites, two on the first tetrahedron and one on the
second. The remainder of the Ce sites are then assigned the valence of $+3$
as illustrated in Fig. \ref{ce7o12bvm1}(d)(i). An alternative arrangement of the
charges is illustrated in Fig. \ref{ce7o12bvm1}(d)(ii). Here, the three $+4$
valences are assigned one each to one of the $i$ sites from the first and
second tetrahedra with the third being assigned to the $S_{6}$ site.
%

\subsection{\label{comparison}Comparison with calculations based on Density Functional Theory}
Most DFT calculations (performed in the supercell of composition Ce$_{32}$O$_{63}$) characterizing the location and character of the excess charge near bulk O vacancies concluded that the charge localizes on two of the nearest neighbour Ce ions \cite{Skorodumova2001,Skorodumova2002,Fabris2005,Kresse2005,Esch2005,Nolan2006a,Andersson2007,Keating2009}. This is what we have called \emph{the standard picture}. 
It appears that this picture was first brought into question by Castleton \textit{et al}. \cite{Castleton2007} in a paper where the question of charge localization was extensively explored. They concluded that it wasn't possible to fully localize the Ce $4f$ electrons on these Ce sites in the DFT+$U$ framework while at the same time preserving a correct description for all the other electrons. We are not aware of any subsequent DFT+$U$ reports directly addressing this question in the conventional supercell. The more recent work used a periodic electrostatic embedded cluster method (PEECM) \cite{burow:174710}. This work found that the charge preferred to localize well away from the O vacancy, being on Ce sites of the third coordination shell. As this shell coincided with the boundary of the quantum mechanical part of the cluster used in the calculation, the authors raised the possibility that the result may be an artifact of the method. 

We also note that Burow \textit{et al}. \cite{burow:174710} obtained three different Ce-Ce distances of $4.07$, $4.12$ and $4.20 \text{\AA}$ in the first coordination shell. These distances are shorter than those reported in Table \ref{shells01} for the vacancy cluster of Ce$_{11}$O$_{20}$. However, since the structures in question are different, one should be cautious about such comparisons.

Our results are also in agreement with some recent reports on the location and localization of the excess electrons for surface and subsurface O 
vacancies \cite{ganduglia-pirovano:026101,li:193401}.
 These authors argued that the localization of charge on next nearest neighbour Ce sites was mainly controlled by the lattice relaxation due to electrostatics as reflected in the lowering of the Madelung potential. However, it should be remarked that a bulk O vacancy is in a different coordination environment relative to either a surface or subsurface O vacancy. It is therefore possible that the comparison between our results for bulk O vacancies and those reported for surface/subsurface O vacancies may not be simple. Indeed, some authors have reported differences in the location of the charge between surface/subsurface and bulk O vacancies \cite{burow:174710}. Nevertheless, it is interesting to note that, until these more recent results, the standard picture was considered applicable to surface and subsurface O vacancies as well \cite{Esch2005,Nolan2006a,Torbrugge2007}.
\section{\label{predict}Predicted charge distribution in Ce$_{6}$O$_{11}$}
\begin{figure}
[ptb]
\begin{center}
\includegraphics[width=0.4\textwidth]
{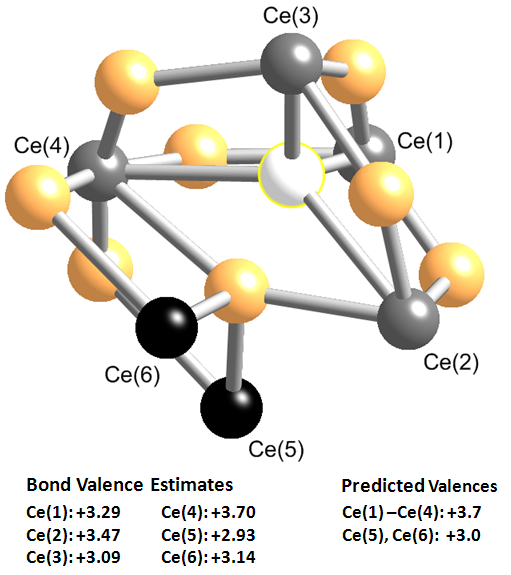}%
\caption{The proposed vacancy cluster of
Ce$_{6}$O$_{11}$ consisting of the six Ce sites as shown. The sites Ce(1) -
Ce(4) are in the first coordination shell of the O vacancy and are
coordinated to $7$ O atoms. The Ce(5) and Ce(6) sites are in the second
coordination shell and have the full $8$-coordination of the Ce site in CeO$%
_{2}$. Bond valence sums are listed for each of these Ce sites as estimated
from the positional parameters of Pr$_{6}$O$_{11}$ and taken from \textbf{I}%
. We also show the site valences we predict from a generalization of the
bond valence results of Ce$_{7}$O$_{12}$ and Ce$_{11}$O$_{20}$.}%
\label{ce6o11vacancy}%
\end{center}
\end{figure}

We make a few remarks about a possible charge distribution around the O
vacancies in this crystal whose unit cell is shown in Fig. \ref%
{ce6o11unitcell}. Complete crystal structure data is not available but we
make predictions of the charge distribution to be expected in this crystal
from a generalization of the bond valence results of Ce$_{7}$O$_{12}$ and Ce$%
_{11}$O$_{20}$.
\begin{figure}
[ptb]
\begin{center}
\includegraphics[width=0.4\textwidth]
{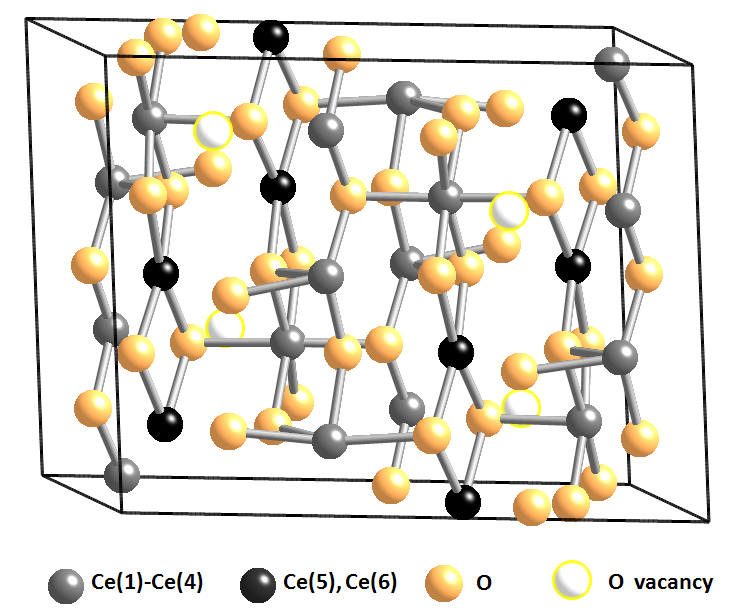}%
\caption{The unit cell of Ce$_{6}$O$%
_{11} $ which consists of four formula units. The Bravais lattice is
monoclinic and of space group $P2_{1}/c$.}%
\label{ce6o11unitcell}%
\end{center}
\end{figure}
%
The proposed vacancy cluster for studying the charge distribution in the
local environment of the oxygen vacancy in this crystal is shown in Fig. \ref%
{ce6o11vacancy}. We did not find a full
crystallographic characterization of this crystal and therefore it is not
possible to perform accurate bond valence calculations for it. The results
reported in \textbf{I} for this crystal were based on the positional
parameters of Pr$_{6}$O$_{11}$. If one were to accept these results (Table I
of \textbf{I), }then since Ce(1), Ce(2), Ce(3) and Ce(4) coordinate the O
vacancy with Ce(5) and Ce(6) in the second coordination shell for the Ce
sublattice, we see that the charge distribution does not quite follow the
pattern one would expect from the results obtained for Ce$_{11}$O$_{20}$ and
Ce$_{7}$O$_{12}$. The overall charge in the vacancy cluster is larger by
about $\sim -1.2$ compared to what one would expect starting from mixed
valence CeO$_{2}$. This discrepancy is twice what we found in the case of Ce$%
_{11}$O$_{20}$ if the discrepancies are normalized to the total number of Ce
sites in each vacancy cluster. As a result, the three Ce sites Ce(1), Ce(2)
and Ce(3) which one would expect to have valences close to $+3.7$ have much
lower valences. We do not believe that this is the correct charge
distribution in the local environment of the O vacancy for this crystal.
Instead, we prefer to extrapolate the argument which emerges from the bond
valence results of Ce$_{11}$O$_{20}$ and Ce$_{7}$O$_{12}$. Doing so leads to
the prediction that the correct charge distribution for this crystal should
be as illustrated in Fig. \ref{ce6o11vacancy} where all the charge goes to
the two Ce sites in the second coordination shell i.e. Ce(5) and Ce(6).
%
Of the known crystal structures of the reduced higher oxides of ceria, Ce$%
_{6}$O$_{11}$ is the closest to the customary supercell used in DFT in that
it is the least reduced. The prediction we have made here would suggest
that, except in an artificial way, it should not be possible to localize the
electrons from the O vacancy site in the first coordination shell for the
customary supercell used in DFT for studying this problem. An experiment
giving the full crystallographic data for this phase would help resolve this
question.
%

\section{\label{matrix}Delocalisation via  $f$-$p$ hybridisation}      

The possibility of the direct hopping of electrons between $f$-orbitals on
neighbouring Ce sites has been suggested to explain the charge distribution
in reduced ceria phases \cite{Zhang1993}. We briefly explore this suggestion
considering the example of Ce$_{7}$O$_{12}$. There are two immediate
questions one may like to answer about the charge distribution obtained for
this phase from the bond valence calculations. 
First, how does an $f$-electron delocalize among the Ce(2) sites?
 Is this through direct $f$-$f$ hopping or through the more indirect
two-step process of $f$-$p$ hopping which involves an O site bonded to both
Ce sites? Second, why does this delocalization of the charge not extend to
include the Ce(1) which is also in the same first coordination shell of the
O vacancy? We are not in a position to give a detailed answer to these
questions in the present paper but only some preliminary indications based
on Harrison's method of universal parameters \cite{Harrison2004a}. We do not
believe that the delocalization of the $f$-level charge between the Ce(2)
sites is a result of direct $f$-$f$ coupling between these sites.

We now consider the two relevant cases, i.e., direct $f$-$f$ hopping between
neighbouring Ce sites and the indirect $f$-$p$ hopping which involves an O
site between the Ce sites in question. In each of these cases, we consider
two situations: electron hopping between two Ce(2) sites which delocalizes
the charge between these sites and the electron hopping between a Ce(2) and
a Ce(1) site which could also delocalize the charge between these sites.

For the case of direct $f$-$f$ coupling, the relevant distances are:
Ce(1)-Ce(2) - $4.11 \text{\AA}$. These distances are quite comparable and they give the same direct $f$-$%
f $ hopping matrix element, $t_{ff}$, be estimated \cite{Harrison2004a,PhysRevB.28.550} to be about $0.01$ eV. This matrix element is the
most favourable of all the four nonvanishing $f$-$f$ matrix elements and
represents a $\sigma $-$\sigma $ interaction between the two orbitals. This
result would suggest that, if direct $f$-$f$ coupling were the dominant
mechanism of charge delocalization, then charge would delocalize over all
four Ce atoms which are nearest neighbours to the O vacancy.

In the case of $f$-$p$ hopping, we again, consider two cases: hopping along the Ce(1)-O(2)-Ce(2) bonds which we designate `O(2) hopping' and the `O(1) hopping which occurs along Ce(2)-O(1)-Ce(2).  For O(2) hopping, the respective matrix
elements, $t_{fp}$, for electron hopping between Ce(1) and O(2) as well as
O(2) and Ce(2) are \cite{Harrison2004a,PhysRevB.36.2695} $0.67$ and $%
0.46$ eV. These matrix elements refer to a $\sigma $-$\sigma $
interaction between the two orbitals which are the more favourable of the
two nonvanishing matrix elements.
%
Since, for $f$-$p$ hopping to couple $f$ states between two Ce sites, a
two-step process involving the hopping of an $f$-electron from one Ce(2)
site to an O $2p$ level followed by the hopping of an electron from the O $%
2p $ to the $4f$-level of the other Ce(2) site is required, the overall
matrix element for O(2) hopping, $t_{eff}$, becomes $0.15$ eV calculated from Eq. (\ref{hopping}).%
\begin{equation}
t_{eff}=\frac{t_{fp}^{2}}{\varepsilon _{f}-\varepsilon _{p}}  \label{hopping}
\end{equation}%
Where we have assumed that the energy gap between the Ce $4f$- and the O $2p$
levels, $\varepsilon _{f}-\varepsilon _{p}$, is $\sim 2$ eV \cite{Castleton2007}. The corresponding matrix elements for O(1) hopping $0.50$, $0.61$ and $0.15$ eV respectively. We see that the overall matrix element, $t_{eff}$,
is the same in the two cases. Again, this
would suggest that the charge should completely delocalize over all the Ce
sites in the divacancy which contradicts the results of the bond valence
method. Thus, we conclude that the Harrison parameters cannot fully explain
the charge delocalization in the Ce$_{7}$O$_{12}$ divacancy. However, they
appear to exclude the possibility of direct $f$-$f$ hopping since $t_{eff}$
is an order of magnitude larger than $t_{ff}$. These approximate results give some indication of why the charge
delocalization may not extend to include the Ce(1) site. An alternative way
of looking at the indirect $f$-$f$ coupling is to view it as a hybridization
of the $4f$ states at the two Ce(2) sites and the O $2p$ states at the O
atom connecting them through Ce-O bonds. This indirect coupling is analogous
to the superexchange mechanism first discussed by McConnell \cite%
{McConnell1961}.
\section{Conclusions and future directions}
\label{Sec VII}
The goal of this review was to answer the question: when an oxygen atom is removed from bulk cerium oxide where do the two excess electrons go?
The approach taken was to consider high resolution crystal structures of
Ce$_{11}$O$_{20}$ and Ce$_{7}$O$_{12}$ and see how they could be viewed
as ordered arrays of oxygen vacancies in an underlying CeO$_2$ crystal.
The charge distribution in the local environments of the O vacancies can
then be deduced from the bond valence model. 
An important conclusion is that the results are incompatible with
the widely accepted standard picture of charge localization on two
cerium ions next to the vacancy.
Instead, we found that the charge distributes itself 
predominantly in  the second coordination shell
of cerium ions. Furthermore, one excess electron can be delocalised
over more than one cerium ion.

Our conclusions concerning the charge distribution
near oxygen vacancies are significant for several reasons.
First, they contradict many (but not all) atomistic simulations based on density functional theory. Second, the actual charge distribution around
the defect has important implications for the other questions
we posed at the beginning of this review.
The charge distribution has a significant effect on the relative
stability of surface and subsurface vacancies\cite{ganduglia-pirovano:026101}.
Also, if the charge around oxygen surface and subsurface vacancies is not simply
localised on Ce ions next to the vacancy this could change our understanding
of the catalytic activity of these surfaces since it has been claimed or assumed that it is associated with Ce$^{3+}$ ions at the surface.\cite{Campbell2005,Esch2005}
The charge distribution around the vacancies has implications for electronic and
ionic conduction, a subject we have discussed elsewhere\cite{Shoko2009c}.

Since our determination of the charge distribution around vacancies is
indirect it will be important to see whether the same conclusions
are obtained on experiments on these higher order oxides with
complementary probes such as Electron Spin Resonance (ESR), nuclear magnetic
resonance, and photoemission spectroscopies. 
For example, ESR spectra should quite be distinct for
the standard picture of electrons localised onto Ce$^3+$ ions
and our alternative picture of partial delocalisation over the second co-ordination shell and unique charge (and spin) distributions associated
which single vacancies and divacancies.

A variety of spectroscopic probes have shown that the formation of
oxygen vacancies is also associated with next electronic states located
in energy between the valence band of nominally ``oxygen 2p" states and
the ``localised 4f" states.\cite{Mullins1998,Henderson2003}
DFT-based calculations which do not sufficiently
include the effect of electronic correlations
fail to produce these states.\cite{Ganduglia-Pirovano2007} 
Generally the energy of defect states is correlated with the extent of electron
localisation around defects. 
Hence, we suggest that spectroscopic studies of ordered
phases of higher order oxides could
be a fruitful approach to characterising the electronic properties
of defects and complementary to the structural approach we have reviewed.
We expect that the energies of the electronic states associated with each
of the vacancies we have considered will be different and correlated
with the extent of charge delocalisation that we find.
Such experimental results will provide significant constraints on theories.

This review only considered bulk cerium oxides. However, we believe
our approach to characterising electronic properties of oxygen vacancies in
terms of a bond valence sum analysis and the physical insights gained should also be fruitful in the study of surface defects and to other widely studied oxidessuch as those of titanium, hafnium, vanadium, and zirconium.

\section{Acknowledgements}
This work was supported by the Australian Research Council.
E.S. was supported by an 
Australian Commonwealth Government 
International Postgraduate Research
Scholarship and a University of
Queensland International Postgraduate Research Scholarship.
He also received a
School of Mathematics and Physics Postgraduate Travel
Scholarship and a Travel award from the Australian Research Council Nanotechnology Network.
We have benefited from discussions with 
A. Jacko, B.J. Powell, S. Olsen, D. Sholl, J. Reimers, N.S. Hush,  
W. Soboyejo, and G. Palenik.

\bibliography{shoko_jpcm_09}
\end{document}